\documentclass[iop]{emulateapj}

\newcommand{\Ms}{M$_\odot$}

\slugcomment{Accepted by ApJL}
\shorttitle{The Magnetic Field in the Class 0 Protostellar Disk of L1527}
\shortauthors{Segura-Cox et al.}

\begin{document}


\title{The Magnetic Field in the Class 0 Protostellar Disk of L1527}

\author{Dominique M. Segura-Cox\altaffilmark{1}, Leslie W. Looney\altaffilmark{1}, Ian W. Stephens\altaffilmark{1,2}, Manuel Fern\'{a}ndez-L\'{o}pez\altaffilmark{1,3}, Woojin Kwon\altaffilmark{4}, John J. Tobin\altaffilmark{5}, Zhi-Yun Li\altaffilmark{6}, Richard Crutcher\altaffilmark{1} }
\altaffiltext{1}{Department of Astronomy, University of Illinois, Urbana, IL 61801, USA; segurac2@illinois.edu}
\altaffiltext{2}{Institute for Astrophysical Research, Boston University, Boston, MA 02215, USA}
\altaffiltext{3}{Instituto Argentino de Radioastronom\'{i}a, CCT-La Plata (CONICET), C.C.5, 1894, Villa Elisa, Argentina}
\altaffiltext{4}{SRON Netherlands Institute for Space Research, Landleven 12, 9747 AD Groningen, The Netherlands}
\altaffiltext{5}{National Radio Astronomy Observatory, Charlottesville, VA 22903, USA}
\altaffiltext{6}{Astronomy Department, University of Virginia, Charlottesville, VA 22904, USA}
    
\begin{abstract}
We present subarcsecond ($\sim$0.35$^{\prime\prime}$) resolved observations of the 1.3~mm dust polarization
from the edge-on circumstellar disk around the Class 0 protostar L1527.  
The inferred magnetic field is consistent with a dominantly toroidal
morphology; there is no significantly detected vertical poloidal component to which observations
of an edge-on disk are most sensitive.  
This suggests that angular momentum transport in Class 0 protostars (when large amounts of material
are fed down to the disk from the envelope and accreted onto the protostar) is driven mainly by  magnetorotational instability rather than magnetocentrifugal winds at 50 AU scales.
In addition, with the data to date there is an early, tentative trend that
R$>$30 AU disks have so far been found in Class 0 systems with average magnetic fields on the 1000 AU scale strongly misaligned with the rotation axis.  The absence of such a disk in the aligned case could be due to efficient magnetic braking that disrupts disk formation.
If this is the case, this implies that candidate Class 0 disk systems could be identified by the average magnetic field
direction at $\sim$1000 AU spatial scales.
\end{abstract}


\section{Introduction}

Circumstellar disks are a key component of the star formation
process and are fundamental for accretion and
angular momentum distribution during the early phases of star
formation.  Class 0 objects are the youngest and most embedded
protostars, and circumstellar disks form at
this earliest stage of star formation if angular momentum is conserved
during cloud collapse \citep[e.g.,][]{cm81}.  Class 0
disks are extremely obscured by envelopes, which contribute $\gtrsim$90\% of the total emission
\citep{lo00}, making the search
for Class 0 disks challenging.
However, observations of Class 0 disks and their properties are essential to
provide the initial
conditions for mass accretion onto the central protostar and planet
formation.  To date, only a few Class 0 systems have observed disks with clear Keplerian
rotation \citep[e.g.,][]{to12,mu13,cd14};
L1527, VLA 1623, and HH212 have Keplerian disks with R$>$30 AU, sizes larger than magnetic braking models predict.

In addition to the properties of young disks and envelopes, the
morphology and strength of the magnetic field in these systems 
also play an important role in star formation \citep[e.g.,][]{cr12}.
For example, the morphology of the magnetic field in the young disk
provides important clues into angular momentum transport: disk accretion driven by magnetorotational
instabilities \citep[MRI,][]{bh98} favor toroidal fields while angular momentum removal via magnetocentrifugal winds arising from the disk favor poloidal fields
\citep[e.g.,][]{bp82}.  The magnetic field morphology in the envelope
and disk can be inferred; dust grains preferentially align with
their long-axis perpendicular to the magnetic field, causing the
dust emission to be polarized \citep[e.g.,][]{lz07}.  An interferometric
survey of dust polarization around 26 low-mass Class 0/I
protostars has been recently conducted \citep[TADPOL,][]{hu13,hu14}.
For these sources, the average magnetic field
axes are generally misaligned with the rotation axes of the systems
(as proxied by the outflow).  In the case of L1527 and VLA 1623,
TADPOL observations show that the magnetic field lines are perpendicular
to the outflows.  However, the TADPOL results only probe envelope
size scales and do not approach disk scales.  The magnetic field
morphology on smaller scales has been observed in the Class 0
protostars L1157 and IRAS 16293-2422 B.  L1157, whose disk is yet
to be resolved and has R$<$20 AU \citep{to13a}, has
vertical poloidal component magnetic fields aligned with the rotation axis of the
system in the inner envelope \citep{st13}.  IRAS 16293-2422 B---which
is thought to have a face-on disk and hence no clear Keplerian
motion---has resolved observations of the candidate disk with a
polarization pattern indicative of a toroidal magnetic field component
\citep{ro14}, although the face-on geometry makes the detection of
any vertical poloidal component impossible.

In this Letter, we present high-resolution CARMA 1.3~mm dust polarimetric
observations of the Class 0 protostar L1527.  Lower-resolution CARMA
1.3~mm polarimetric observations of L1527 were previously conducted
as a part of the TADPOL survey, probing the magnetic field morphology
on $\sim$1000 AU envelope size scales.  Here we report the magnetic
field morphology on $\sim$50 AU disk size scales.  The data presented
are the first detection of polarized dust emission emanating
directly from a Class 0 Keplerian disk.

\section{Observations}

CARMA 1 mm full-Stokes observations of L1527 were obtained in 6 tracks
of the B array ($\sim$0.35$^{\prime\prime}$
resolution) from 2013 December 9--13 and 15 for a total of 21 hours on-source.
The correlator was set up with a local oscillator frequency of 233.731 GHz and four 500 MHz-wide bands centered at intermediate frequency values of 2.187, 2.740, 4.716, and 5.544 GHz.
We used the MIRIAD software package \citep{sa95} to reduce the
data.
The polarization calibration followed the standard process
for CARMA \citep{hu14}.
The phase and polarization leakage calibrator
for all tracks was 0510+180.  The preferred bandpass calibrator was
3C84, and 3C454.3 was used when 3C84
was unavailable.  For most tracks, the flux calibrator was MWC349
with a calibration accuracy estimation of $\sim$15\%, but only statistical uncertainties are considered in this Letter.  When this calibrator was
not observed, the flux was interpolated from the observations of
MWC349 in other tracks.  
To maximize sensitivity, 
maps of the 4 Stokes parameters (I, Q, U, and V) were
created using natural weighting.  From these, we derive the
polarization position-angle and intensity maps.
We also generated new maps of L1527 data from the TADPOL survey
\citep{hu14}.  The data include 5 tracks between 2011 May and 2013 April in
D and E arrays.

\section{Results}

\begin{figure}[t]
\centering
\includegraphics[width=0.45\textwidth]{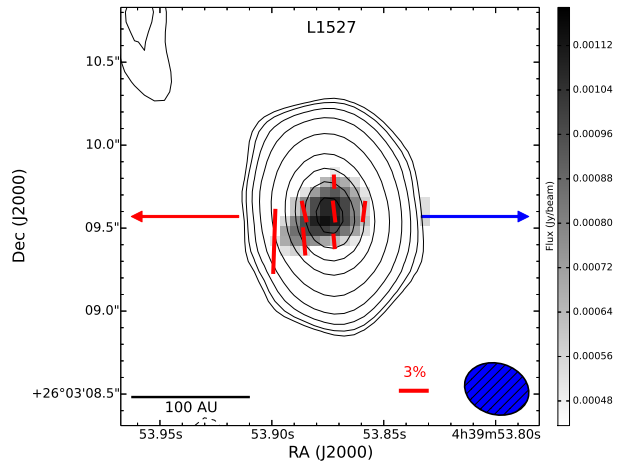}
\caption{Polarimetric map (polarization vectors rotated by 90$^{\circ}$ to show inferred magnetic field orientation) of the L1527 disk from CARMA data with a 0.39$^{\prime\prime}$$\times$0.31$^{\prime\prime}$ beam. Fractional polarization vectors $\geq$3$\sigma$ displayed. Contours are Stokes I data with levels of [-6, -4, -3, 3, 4, 6, 10, 20, 40, 60, 80, 100]$\times$$\sigma$, $\sigma$=0.45~mJy beam$^{-1}$. Grayscale shows the polarized intensity $\geq$3$\sigma$. Outflows in the plane of the sky are marked by red and blue arrows.}
\label{fig1}
\end{figure}

The 1.3~mm dust emission map of L1527 is presented in Figure \ref{fig1} and is consistent with
the known edge-on disk \citep{to13b}.
The disk has been resolved previously at 3.4~mm and 870~$\mu$m \citep{to12} and also has been shown
to have Keplerian motion and a radius of 54 AU \citep{oh14}. 
Our observations are the first resolved detection of the
disk at 1.3~mm.
At a resolution of $\sim$0.35$^{\prime\prime}$ and
a distance of 140 pc \citep{ld07}, our interferometric observations probe L1527
on size scales of $\sim$50 AU.
An elliptical Gaussian fit to the
high-resolution 1.3~mm Stokes I data measures a deconvolved size of
0.53$^{\prime\prime}$$\times$0.23$^{\prime\prime}$ with position angle of 5.2$^{\circ}$ (PA, measured counterclockwise), 
consistent with the deconvolved sizes at 3.4~mm, 870~$\mu$m \citep{to13b}, and 1.3~mm \citep{oh14}.
In addition, our flux density at 1.3~mm (139$\pm$4~mJy) is consistent with 
detection of the L1527 disk seen
in \citet{to13b}.
Using the measured fluxes at 3.4~mm and 870~$\mu$m and the derived $\beta$=0 from \citet{to13b},
we can estimate the expected disk emission at 1.3~mm (116~mJy and 96~mJy, respectively), 
which is congruent with our measured 1.3~mm fluxes when taking account a 20\% extrapolating 
and amplitude uncertainty.
Based on this evidence, our observations are dominated by disk emission of the L1527 system,
with little contamination from the large-scale envelope emission. 

We detect dust polarization of the young disk over 2 synthesized
beams with an average polarization
of 2.5\%$\pm$0.6\% and a position angle of 5$^{\circ}$$\pm$5$^{\circ}$ measured 
counterclockwise from north,
aligning well with the Stokes I elliptical Gaussian fitted position angle of 5.2$^{\circ}$$\pm$0.4$^{\circ}$. 
The inferred magnetic field (with polarization vectors rotated by 90$^{\circ}$) is shown in Figure \ref{fig1}.  The morphology of the inferred field is parallel to the disk axis, as is expected 
from an edge-on toroidal field---uniform and aligned with the disk.
We compare a uniform field at a 5$^\circ$ PA (the same as the dust emission)
to the data and find a reduced $\chi^2<1$.
The polarization fraction of the circumstellar disk of L1527 is
larger than the 1.4\% polarization fraction found in the face-on
candidate disk of IRAS 16293-2422 B \citep{ro14}, although the lower polarization
fraction of IRAS 16293-2422 B may be due in part to beam-averaging;
due to orientation, an edge on toroidal field is less
beam-averaged as the vectors are more uniform.
Our polarization percentage is similar to the theoretically predicted
2-3\% polarization
fraction found in simulations of magnetized disks \citep{cl07}.
On the other
hand, observations of the disks of older T~Tauri systems have much
lower polarization percentages $<$1\% \citep{hg09,hg13}, which may
be an outcome of dust processing or de-alignment mechanisms during
disk evolution \citep{st14}.

\begin{figure*}
        \centering
                \includegraphics[width=0.3\textwidth]{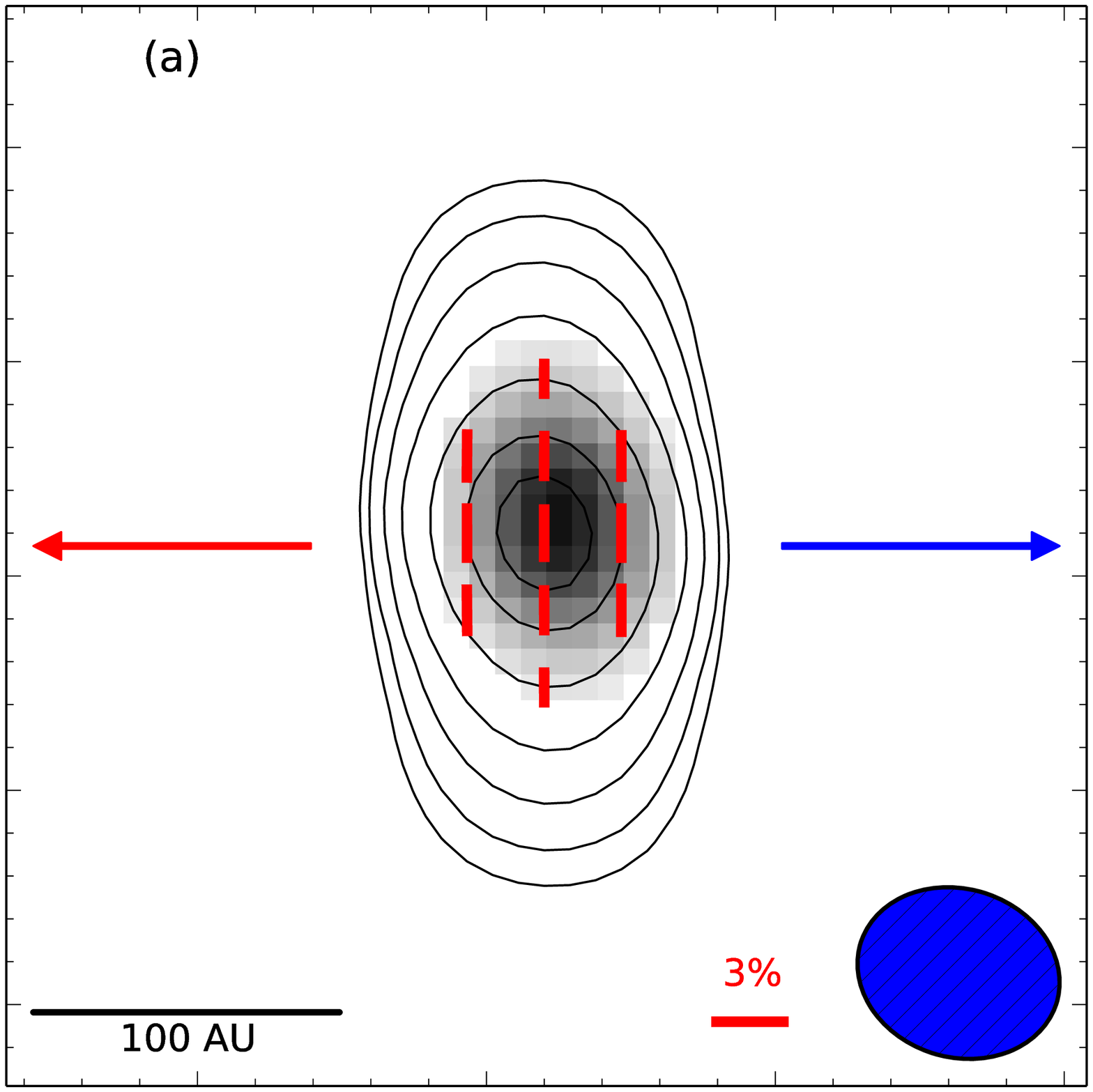}
                \label{fig2a}
        \hspace{-0.25cm}
        \vspace{-0.1cm}
                \includegraphics[width=0.3\textwidth]{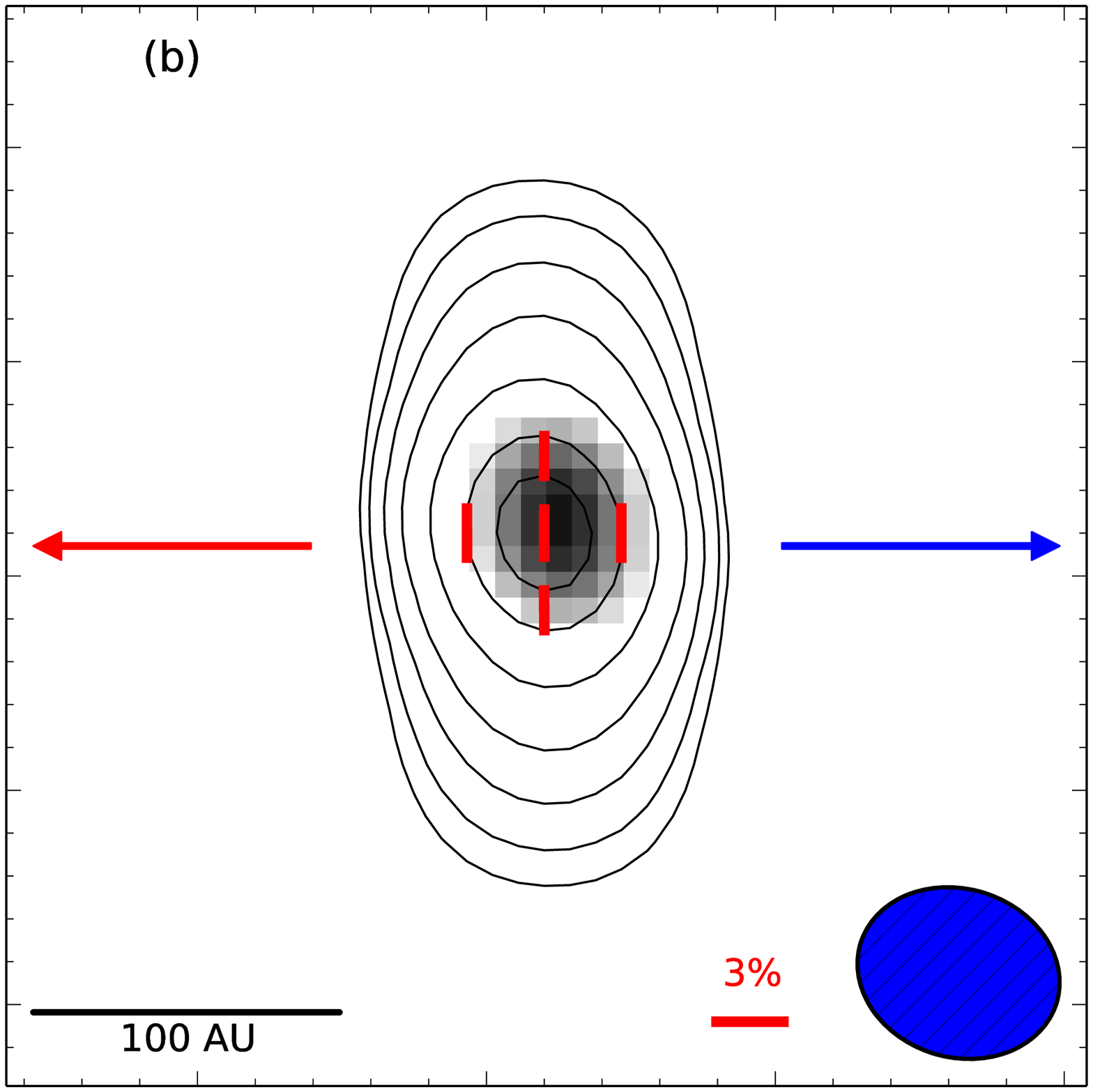}
                \label{fig2b}
                         \hspace{-0.25cm}
                \includegraphics[width=0.3\textwidth]{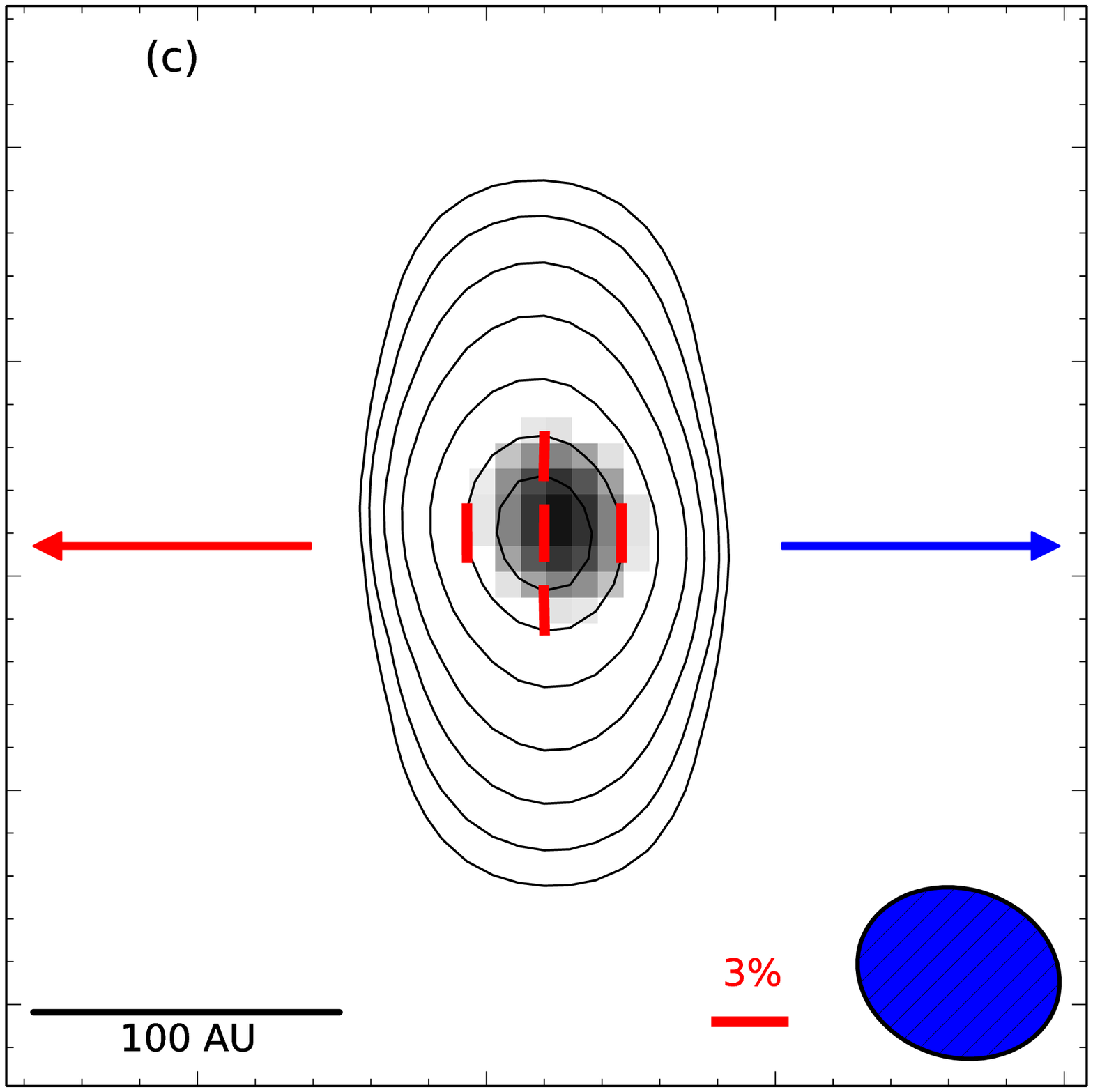}
                \label{fig2c} 
        \hspace{-0.25cm}
                \includegraphics[width=0.3\textwidth]{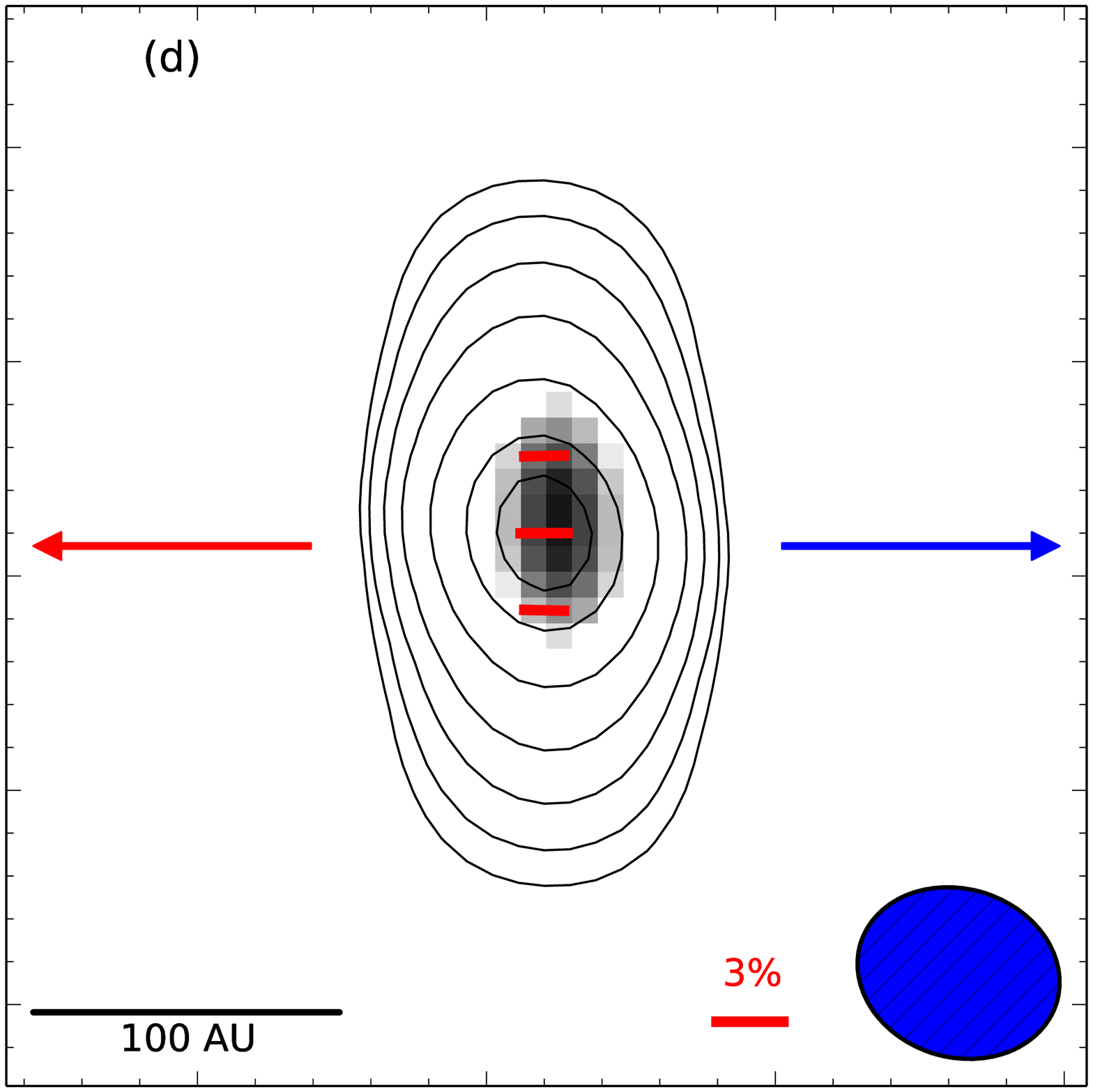}
                \label{fig2d}       
                        \hspace{-0.25cm}       
                \includegraphics[width=0.3\textwidth]{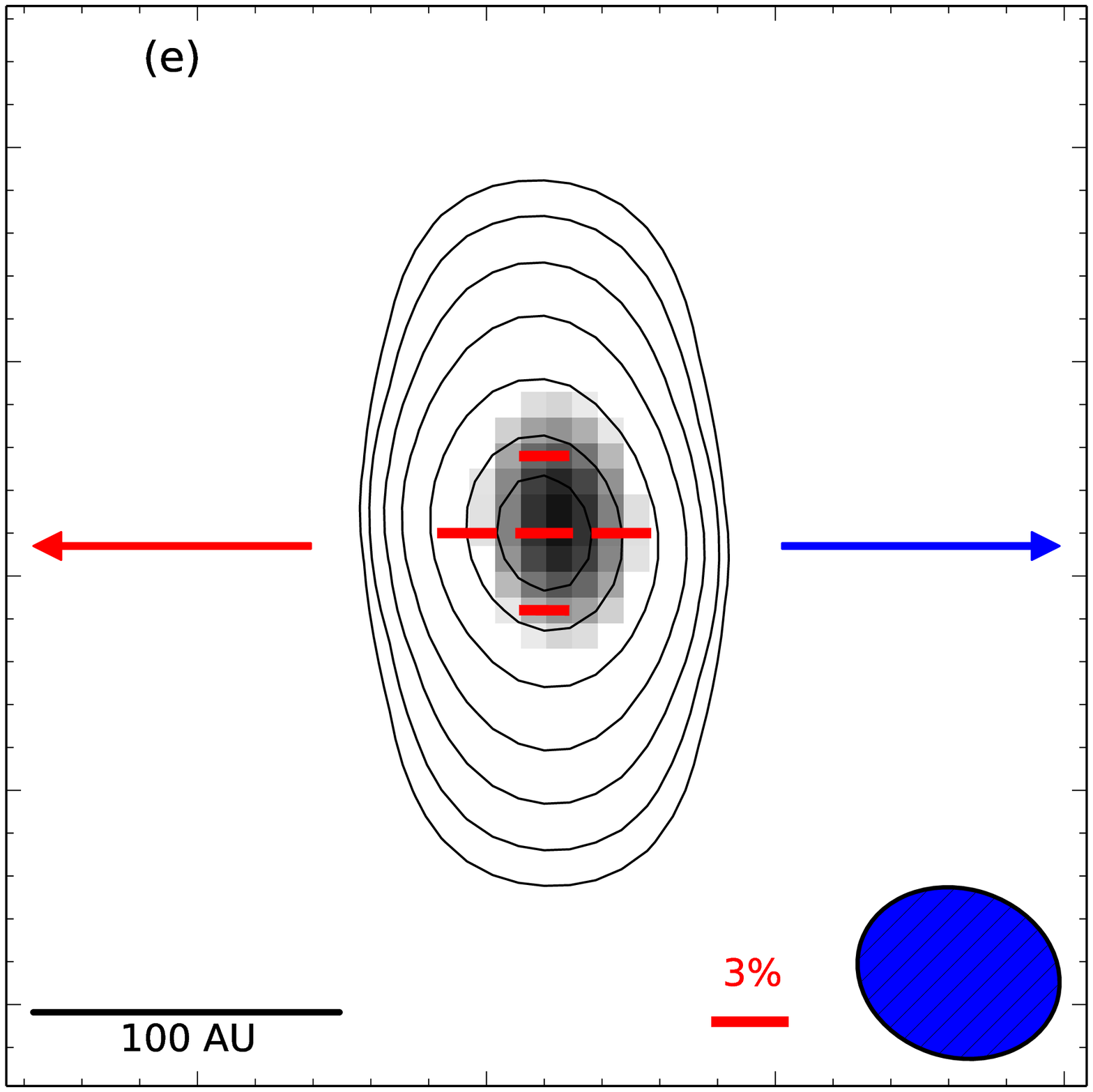}
                \label{fig2e}
       \hspace{-0.25cm}
                \includegraphics[width=0.3\textwidth]{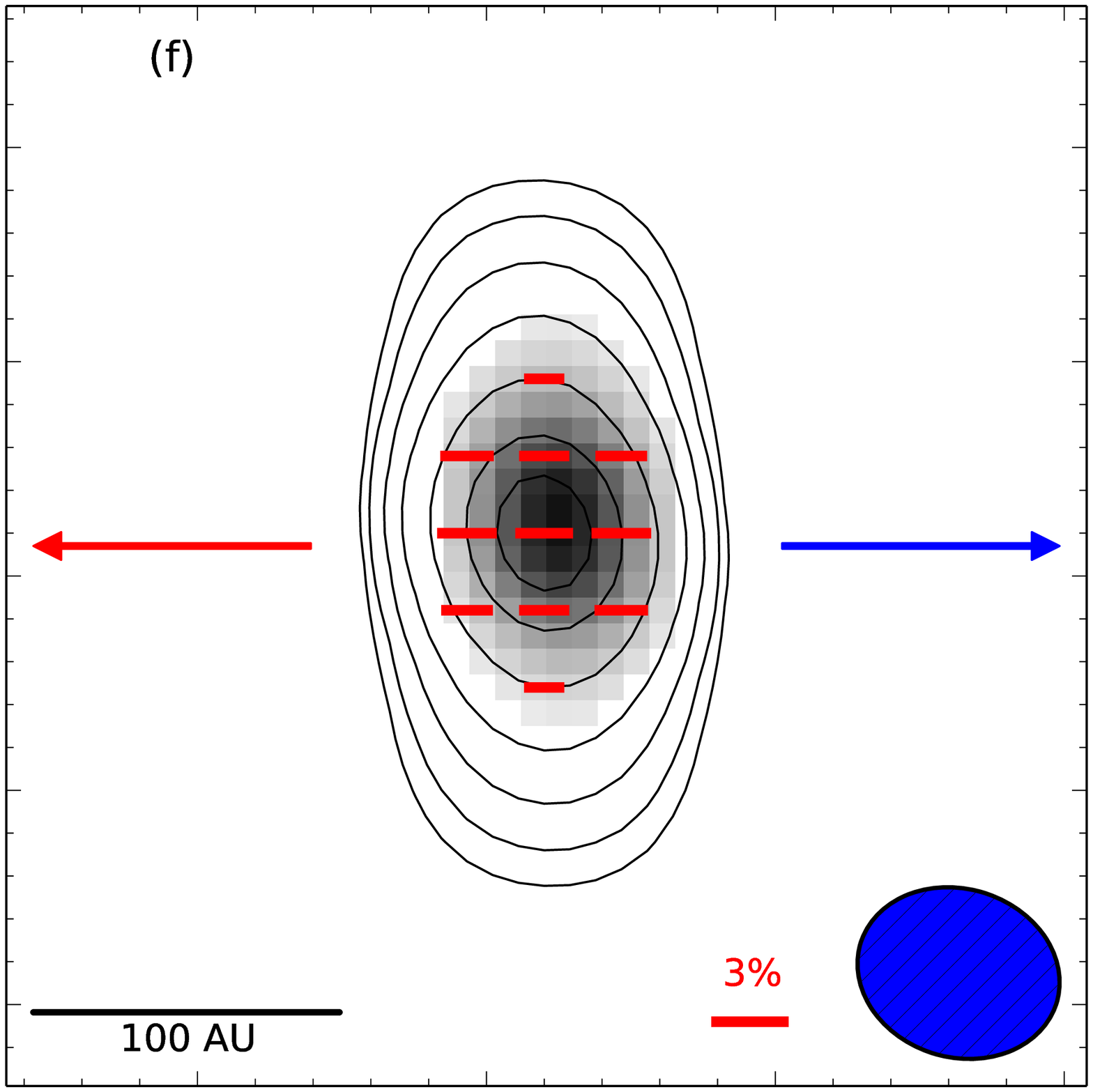}
                \label{fig2f}
        \caption{Synthetic maps of the L1527 disk magnetic field morphology. Contours, grayscale, and vectors are the same as Figure 1.  (a) Toroidal field only, (b) 70\% toroidal/30\% vertical poloidal field, (c)  60\%/40\%, (d) 50\%/50\%, (e) 40\%/60\%, (f) vertical poloidal field only.  
        }\label{fig2}
\end{figure*}

A uniform field in the plane of the disk is physically unlikely for a rapidly-rotating,
Keplerian disk system.  One expects either poloidal, toroidal, or a combination of the two
in such a disk \citep[e.g.][]{bh98,kp00}.
For an edge-on disk, observations are most sensitive to vertical poloidal field
components because they are expected to vertically thread the disk and thus lay roughly in the plane
of the sky.  However, our data does not exhibit any obvious poloidal
morphology which would be perpendicular to the disk.
To show that a toroidally dominant morphology is consistent with our observations,
we compare to a purely toroidal, a purely vertical poloidal, and combinations of toroidal and vertically poloidal toy models (see caption of Figure \ref{fig2} for details).
We used the best-fit disk parameters of \citet{to13b}: disk inclination angle of 85$^{\circ}$, 0$^{\circ}$ PA, 
M$_{\mathrm{disk}}=0.0075$ \Ms, R$_{\mathrm{inner}}=0.1$ AU,
R$_{\mathrm{outer}}=125$ AU, and stellar and accretion luminosity of 2.75 L$_{\odot}$.
The temperature distribution was calculated using the 
Monte-Carlo radiative code RADMC-3D 
\citep{dd04}.
The Stokes I, Q, and U maps were numerically solved using dust radiative transfer in the
disk along the line of sight with an assumed constant polarization fraction.
The magnetic field vectors
at each integral element have been tilted and rotated based on the
inclination and position angles of the disk model.  Finally, all maps
are convolved with the synthesized beam from the polarization
observations \citep[for more details see][]{st14}.  As
shown in Figure \ref{fig2}, the detected polarization is characterized
well by our toroidally dominant toy models with up to 40\% poloidal field (Figure \ref{fig2}a, \ref{fig2}b, \ref{fig2}c), with a reduced $\chi^2\sim$1.8 
for all three cases as fit to the morphology in the image plane with the vectors shown in Figure \ref{fig1}.  The scenarios where the vertical poloidal component is
 equally strong as the toroidal component (Figure \ref{fig2}d) or the vertical poloidal component is 
 dominant (Figure \ref{fig2}e, \ref{fig2}f) do not reproduce the observed magnetic field morphology. The 
 purely toroidal simple model exhibits more extensive
polarization than the observations, which is likely due to our
assumption of constant polarization in the disk.

\begin{table*}[t!]
\caption{B array and TADPOL Results}
        \vspace{-0.5cm}
\begin{center}
\begin{tabular}{lccccc}
\hline
Data Set & Stokes I Flux & Polarized Flux & $\bar{P}_{\%}$  & PA &  Beam \\
& (mJy) & (mJy) & (\%)  & ($^{\circ}$) &  ($^{\prime\prime}$)\\
\hline
B array & 139$\pm$4 & 1.1$\pm$0.2 & 2.5$\pm$0.6 & 5$\pm$5  & 0.39$\times$0.31 \\
TADPOL & 188$\pm$5 & 2.3$\pm$0.3  & 2.8$\pm$0.7 & 1$\pm$5 & 2.63$\times$2.26 \\
\hline
\end{tabular}\\
\end{center}
\tablecomments{Uncertainties are statistical.  Results were found using data $>$3$\sigma$.   Position angles are measured counterclockwise.  Stokes I flux is measured across the entire disk or inner envelope; polarized flux and $\bar{P}_{\%}$ are measured in the polarized region only.}
\label{tab1} 
\end{table*}

\section{DISCUSSION}

In quiescent (non-turbulent) systems with aligned magnetic field and disk rotation axes, 
magnetic braking can have a significant effect on the infalling material in the ideal MHD limit,
removing angular momentum \citep{ml08,hf08}, and 
suppressing growth of the early circumstellar disk by allowing larger accretion rates \citep{li11}.
Magnetic braking can be so effective in Class 0 sources that rotationally supported disks are 
limited to R$<$10 AU \citep[e.g.,][]{db10}, only reaching R$\sim$100 AU at the end of the main 
mass accretion phase when the envelope is less massive and magnetic braking becomes 
inefficient \citep[e.g.,][]{dbk12,ml09,ma11}.  Conversely to this prediction, Keplerian 
disks have been detected in Class 0 sources with sizes larger than expected from magnetic braking models.  L1527 and VLA 1623 have disk sizes of R$\sim$54 AU and R$\sim$189 AU respectively \citep{oh14,mu13}, and HH212 has a disk of R$>$30 AU \citep{cd14}.

There are large disks in some young systems,
suggesting that significant magnetic braking has not happened, has
already occurred, or the magnetic field has diffused to the point
where a R$>$10 AU  disk could form.  On the other hand, similar high
resolution observations of the Class 0 protostar L1157 have not
detected a circumstellar disk down to spatial resolutions of $\sim$15
AU \citep{to13a}.  This result suggests that magnetic braking may
have been more significant in L1157 than L1527.
Of course, there are differences in age since L1527 is an older source
and could have been classified as a 
Class I source, were it not viewed edge-on \citep{to08}.
What is clear is that some Class 0 sources have R$>$10 AU 
circumstellar disks and others do not.
Such differences in disk size could be a consequence of misalignment
between the magnetic field and rotation axis, which modifies the strength of
magnetic braking  \citep{hc09,jo12,li13,ku13}.

\begin{figure}[t!]
\centering
\includegraphics[width=0.45\textwidth]{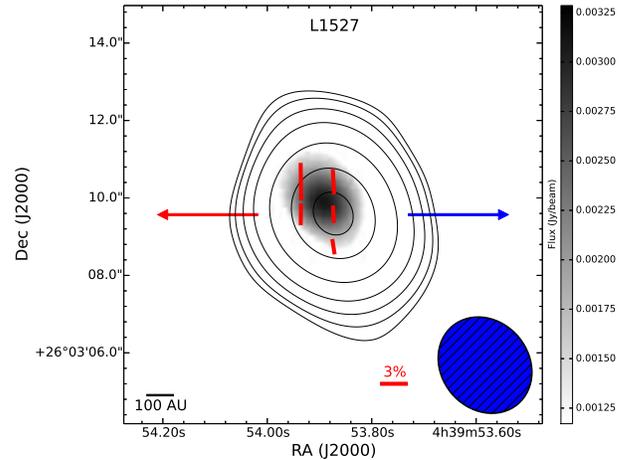}
\caption{Polarimetric map of the L1527 inner envelope from the CARMA TADPOL data with a 2.63$^{\prime\prime}$$\times$2.26$^{\prime\prime}$ beam. Contours are Stokes I  data with levels of [-6, -4, -3, 3, 4, 6, 10, 20, 40, 60, 80, 100]$\times$$\sigma$, $\sigma$=2.34~mJy beam$^{-1}$. Grayscale and vectors are the same as Figure 1.}
\label{fig3}
\end{figure}

To better understand the role of the magnetic field in the early disk and envelope,
we can compare the magnetic field of L1527 presented here with the larger-scale magnetic field
detected with TADPOL (Figure \ref{fig3} and Table \ref{tab1}). 
When comparing the polarization at 1000 AU and 50 AU, the
two scales have the same average field angle: perpendicular to the outflow and
well aligned with the disk plane.
The higher resolution observations have less than half of the polarized
emission, which suggests we are resolving out large-scale emission. 
The projected field morphology on the 1000 AU scale is consistent with the view that the initial magnetic field on this scale is greatly misaligned with respect to the rotation axis, although it is also possible that the field on this scale is already modified by the collapsing and rotating motions in the envelope. The magnetic fields on even larger scales are expected to be affected less by rotation and collapse, and are more likely to keep their initial configuration. The fields are better traced by single dish observations using SHARP on CSO \citet[]{da11} and SCUPOL on JCMT \citep[Figure 17 of][]{hu14,ma09}.  These single dish data are modeled by \citet[]{da14} together with CARMA data; we refer the reader to that paper for a detailed discussion of the magnetic field on large-scales. In any case, on the small scale of the disk, the available data are consistent with the field being predominantly toroidal, and such a
toroidally dominant disk magnetic field is also consistent with the magnetorotational instability
\citep[e.g.,][]{bh98} driving accretion during
the main accretion phase.
We do not detect significant vertical poloidal component
fields that are needed to launch magnetocentrifugal winds; such winds are probably not the dominant driver of angular momentum transport
during the main accretion phase at the $\sim$50 AU size scale.
On the other hand, a disk wind would likely be launched from the disk upper layers that
are not well traced by our observations, which are most sensitive to the dust in the midplane.
 If the poloidal field is somehow limited to the surface layers, then our observations would be less constraining on the existence or absence of a disk wind.
 
\begin{table*}[t!]
\caption{Class 0 Magnetic Field Morphologies and Candidate Disks}
        \vspace{-0.5cm}
\begin{center}
\begin{tabular}{lccccccc}
\hline
Source & $\alpha$ & $\delta$ & Mag-Rot-Axis$^{a}$ & Candidate & Known & References \\ 
 & (J2000) & (J2000) & & Disk?$^{b}$ & Keplerian? & \\
\hline
L1527 & 04:39:53.9 & 26:03:09.6 & Perpendicular & Yes & Yes & 1,2 \\
IRAS 16293-2422 B  & 16:23:22.9 & -24:28:35.7 & Perpendicular & Yes & No & 3,4 \\
VLA 1623 & 16:26:26.4 & -24:24:30.5 & Perpendicular & Yes & Yes & 5,2 \\
L1157 & 20:39:06.2 & 68:02:15.8 & Parallel & No & No & 6,2 \\
\hline
\end{tabular}
\end{center}
\tablecomments{$^{a}$ Orientation of the magnetic field compared
to the rotation axis (estimated from the outflow for L1157).  $^{b}$ Does the source
have a candidate disk of R$>$30 AU?}
\tablerefs{(1) \citet{to12}; (2) \citet{hu14}; 
(3) \citet{za13}; (4) \citet{ro14}; (5) \citet{ms13}; (6) \citet{to13a}} 
\label{tab2}
\end{table*}

With dust polarization observations and high-resolution searches for disks, we can compare the magnetic field orientations and
morphologies with disk properties.
L1527, VLA 1623, and L1157 
have all been observed with CARMA dust polarization
at 500 AU or better resolution  \citep{hu14}.
L1527 and VLA 1623, the first two Class 0 systems
with known Keplerian disks, have average magnetic fields 
perpendicular to the rotation axes (inferred from the outflow direction).  In contrast,
L1157, a system with a disk R$<$20 AU in size, has
an inferred average magnetic field
parallel to the rotation axis.  Although we have few examples so far (Table \ref{tab2}), 
this observational, tentative trend is intriguing and suggested from theory \citep[e.g.,][]{jo12};
the magnetic field morphology at the earliest stages of collapse
may play an important role in the formation of the earliest disk, 
with strongly misaligned magnetic fields and rotation axes producing R$>$10 AU disks at early times. 
Clearly more objects are necessary to better
establish this relationship.

We therefore suggest that the morphology of the magnetic field in the inner
envelope ($\sim$1000 AU, TADPOL scales) 
could be another method to help identify candidate Class 0 disk sources
with R$>$10 AU.
Sources with R$>$10 AU disks may have a projected magnetic field on the 1000 AU scale perpendicular to the
rotation axis, and the magnetic field would appear uniform for edge-on disk cases like L1527.
At more oblique viewing angles at high resolution, a purely toroidal field
would be observed as two maxima of fractional polarization along the axis
of rotation on either side of the protostar, with the magnetic field oriented perpendicular to the
outflow axis within the maxima regions \citep[e.g., see Figure 1 in][]{hg13}.  
A purely toroidal disk field observed directly down the rotation axis
will appear as a pattern of concentric circles
\citep[e.g.,][]{pa12}.
Sources with R$<$10 AU disks may have dominant vertical poloidal component fields misaligned with the axis of rotation,
as is the case with L1157
with a characteristic hourglass pinch near the protostar.  
Such a poloidal morphology observed at viewing roughly perpendicular to the rotation axis of the system can appear either
symmetric or asymmetric on either side of the pinch \citep{ka12}, especially for more oblique viewing angles: L1157 has a slight asymmetry.
When a purely poloidal field is observed directly
down the rotation axis, the poloidal pinch is not observed,
but rather has a convergent, spoke-like morphology.
In the extreme cases of purely toroidal and purely poloidal magnetic
field components, the observed morphology alone can be used to
distinguish between the two cases and point towards R$>$10 AU candidate Class
0 disks.

\section{CONCLUSIONS}

L1527 is the first Class 0 protostar with a known Keplerian disk
and direct detection of linearly polarized dust emission from the
circumstellar disk, indicating magnetic fields are aligned perpendicular to the rotation axis of
the disk.  The magnetic field is consistent with toroidally dominant field
lines.  It may be that L1527's large disk arises from the strongly misaligned rotation
axis and magnetic field on large scales, while aligned rotation axes and
magnetic fields inhibit disk formation on R$>$10 AU scales.
The toroidally dominant field morphology favors the magnetorotational instability
\citep{bh98} as the dominant angular momentum transport process in Class 0 circumstellar disks.

L1527 is one of two Class 0 sources (with VLA 1623) where both magnetic fields and Keplerian disks have
been detected.  Both of these sources have
perpendicular magnetic fields and rotation axes  \citep{ms13,to12,hu13}
on 1000 AU scales. The alternative case is where the magnetic field and
rotation axes are parallel on envelope scales, such as the Class 0 source
L1157 with no disk detected down to 20 AU \citep{to13a}.  It is
possible that aligned magnetic fields may have braked rotation so
efficiently as to inhibit the disk formation and growth at early
times.  
The tentative trend of misaligned magnetic field and rotation
 axes in Class 0 systems with disks is suggestive and expected from theory, requiring follow-up to make
 hard conclusions about Class 0 disk formation.


\acknowledgments
We thank Chat Hull and Dick Plambeck for assistance with data reduction.  This research made use of APLpy, an open-source plotting package for Python hosted at http://aplpy.github.com.

Support for CARMA construction was derived from the states of California, Illinois, and Maryland, the James S. McDonnell Foundation, the Gordon and Betty Moore Foundation, the Kenneth T. and Eileen L. Norris Foundation, the University of Chicago, the Associates of the California Institute of Technology, and the NSF. Ongoing CARMA development and operations are supported by the NSF under a cooperative agreement and by the CARMA partner universities.

J. Tobin acknowledges support provided by NASA through Hubble Fellowship
grant \#HST-HF-51300.01-A awarded by the STScI, which is
operated by AURA,
Inc., for NASA, under contract NAS 5-26555.  The NRAO is a facility of the NSF
operated under cooperative agreement by Associated Universities, Inc.

Z.-Y. Li is supported in part by NASA 14AB38G and NSF 1313083. 

\clearpage

\end{document}